\documentclass[useAMS,usenatbib]{mn2e}
\usepackage{graphicx}

\title[Oscillations of Thick Discs Around BHs] {Oscillations of Thick
  Accretion Discs Around Black Holes} \author[E. Rubio-Herrera \&
  W. H. Lee] {Eduardo Rubio-Herrera \thanks{E--mail:
  eduardo@astroscu.unam.mx} and William H.~ Lee \\ Instituto de
  Astronom\'{\i}a, UNAM, Apartado Postal 70--264 C.P. 04510 M\'exico
  D.F. M\'exico. \\} \date{Released 2004 Xxxxx XX}

\pagerange{\pageref{firstpage}--\pageref{lastpage}} \pubyear{2004}

\def\LaTeX{L\kern-.36em\raise.3ex\hbox{a}\kern-.15em
    T\kern-.1667em\lower.7ex\hbox{E}\kern-.125emX}

\begin{document}

\label{firstpage}

\maketitle

\begin{abstract}
We present a numerical study of the response of a thick accretion disc
to a localized, external perturbation with the aim of exciting
internal modes of oscillation. We find that the perturbations
efficiently excite global modes recently identified as acoustic
p--modes, and closely related to the epicyclic oscillations of test
particles.  The two strongest modes occur at eigenfrequencies which
are in a 3:2 ratio. We have assumed a constant specific angular
momentum distribution within the disc.  Our models are in principle
scale--free and can be used to simulate accretion tori around stellar
or super massive black holes.
\end{abstract}


\section{Introduction}

Accretion discs around compact objects (black holes and neutron stars)
are natural laboratories for the study of strong gravitational fields
and their effects (such as Lense--Thirring precession). From the physics
that lies behind these accretion processes and the accompanying
radiation, it is in principle possible to determine important
parameters of the central object, such as mass and spin for the black
hole case \citep{ab2,wa2}, and to place constraints on the equation of
state of dense matter in the case of neutron stars \citep{kl}.

It has also been pointed out that these objects can be potential
sources of gravitational waves, arising from different processes, such
as nonaxisymmetric instabilities \citep{zu}, nonaxisymmetric tori
powered by the spin energy of the black hole \citep{vp}, magnetized
thick accretion discs \citep{mi}, and global oscillations which induce
a time--varying mass quadrupole \citep{za}. The energy released due to
accretion from such systems, if dynamically stable, has also been
proposed as a mechanism for the production of cosmological gamma--ray
bursts. In this case, super--Eddington accretion rates in flows cooled
by neutrino emission feed the central black hole while powering a
relativistic outflow, which produces the GRB. Such a disc possibly
forms following core collapse in a massive rotating star \citep{wo} or
after the merger of two compact objects in a tight binary \citep{npp}.
One mechanism that could potentially produce the collapse of such a
disc onto the black hole is the runaway radial instability.  This was
identified by \citet{ab4} and appears in discs with constant angular
momentum distributions which overflow their Roche lobes at high rates,
and thus increase the mass of the black hole substantially. This
alteration of the potential is a runaway process, and can destroy the
entire disk within a few dynamical times. However, studies of tidal
disruption and torus formation from compact binary mergers
\citep{rj99,l01} show that the distribution of angular momentum in the
disc is far from being constant, and this can in fact suppress the
instability \citep{dm97}.

In the context of low--mass X--ray binaries, observations performed
with RXTE \citep{vdk00} have shown that there are millisecond
oscillations in systems containing neutron stars or black holes
surrounded by an accretion disc. An important result is that in at
least four black hole sources (H1743-322, GRO J1655-40, XTE J1550-564,
GRS915+105), two apparently stable peaks in the power spectrum
appear at frequencies in the hHz range in a 3:2 ratio. A similar
result was found by \citet{ab3} for the neutron star source Sco
X--1. This lends support to the resonance model originally proposed by
\citet{ab2}, and further developed in terms of parametric resonance in
a thin disc by \citet{ab} and \citet{rebusco04}. In this
interpretation, the frequencies reflect epicyclic motion of perturbed
flow lines in the accretion disc, or combinations between these and a
fixed, perturbation frequency \citep{kakls04,lee2}, perhaps due to the stellar
spin in neutron star sources \citep{wi}. Pressure coupling allows
resonances to occur and manifest themselves in the X--ray lightcurve.

Recently, \citet{za} have shown that an extended torus can respond to
external perturbations in a global fashion, in a series of modes whose
frequencies follow the sequence 2:3:4:... . These are attributed to
acoustic p--modes within the torus, excited by an impulsive
perturbation. Follow--up analytical and numerical work
\citep{rez03a,mry04} has extended these results, by calculating the
set of corresponding eigenfrequencies (in height--integrated discs)
and also by investigating the effects of different background metrics
(e.g., Schwarzschild vs. Kerr for a rotating hole). In all cases where
a numerical experiment was carried out, the perturbation was impulsive
and global, affecting the entire torus. This idea has now been
advanced as an explanation for the kHz QPOs in low--mass X--ray
binaries containing the black hole candidates where the 3:2 frequency
ratios have been reported, as mentioned above \citep{rez03b}

In this Letter we show that it is possible to excite these modes in a
thick accretion disc for a perturbation that is local, only affecting
a small portion of the disc strongly. A frequency ratio of 3:2 for the
two strongest modes is apparent. The lower frequency itself is related
to the radial epicyclic value for test particles in circular motion,
shifted to a lower frequency because of the finite extent of the
torus. The second frequency maintains a 3:2 relation with the first.

\section{Initial conditions and numerical method}

\subsection{Hydrostatic equilibrium for a thick torus}

We construct tori which are in hydrostatic equilibrium, have low mass,
$m < M_{BH}$, and a radial extension $L \sim R$, where $R$ is the
distance separating the torus from the central mass, $M_{BH}$. They
are thick in the sense that their vertical extent, $H$, is comparable
to $L$. We neglect the self-gravity of the torus, and additionally
assume azimuthal symmetry.  A polytropic equation of state,
$P=K\rho^\gamma$, has been used for the construction of initial
conditions. 
Integrating the equations of hydrodynamics, it is possible
to write:
\begin{equation}
\frac{\gamma}{\gamma -1}\frac{P}{\rho}=\Phi_{e}+\Phi_{0}=const,
\end{equation} 
where $\Phi_{e}$ is the effective potential. Here $\Phi_{0}$ can be
interpreted as a filling factor of the effective potential
well. Through its variation tori of different sizes are constructed
(Fig. 1 shows their cross sections over one half the $r-z$
~plane). The gravitational potential of the central mass is computed
with the pseudo--Newtonian expression of \citet{pw}:
\begin{equation}
\Phi_{PN} = \frac{-GM_{BH}}{R-r_{g}}              
\end{equation}
which describes the behaviour of a test particle in a strong
gravitational field and, importantly for our purposes, reproduces the
existence and positions of the marginally stable and marginally bound
orbits in General Relativity ($r_{g}=2GM_{BH}/c^{2}$ denotes the
gravitational radius throughout).  With this potential, and using a
constant distribution of specific angular momentum, $l(r)=cst$, the
effective potential is:
\begin{equation}
\Phi_{e} = \frac{-GM_{BH}}{R-r_{g}}+\int \frac{l(r')^2}{2r'^3}dr.
\end{equation}
Finally, we note that with this potential, the frequency of small
radial oscillations for a perturbed circular orbit (i.e., the radial
epicyclic frequency) is given by
\begin{equation}
\kappa=\frac{1}{2\pi}\left[\frac{GM(r-3r_{g})}{r(r-r_{g})^{3}}\right]^{1/2}.
\label{repi}
\end{equation}
\begin{figure}
\includegraphics[width=84mm]{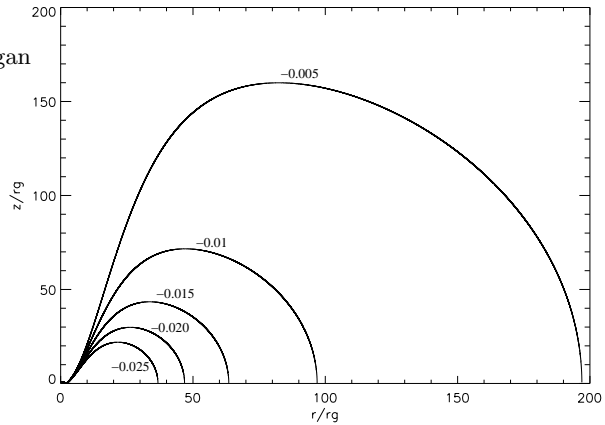}
\caption{Surfaces of zero density and pressure, in meridional cross
section, for various tori constructed according to the procedure
described in the text. Each curve corresponds to different values of
the filling factor $\Phi_0$ in units of $c^2/2$.}
\label{fig1}
\end{figure}

\subsection{Numerical method}

For our simulations, we have used Smooth Particle Hydrodynamics (SPH,
see \citet{mo} for a review), in a two--dimensional version using
cylindrical coordinates. We refer the reader to \citet{lee3} for
details of the implementation. There is no physical viscosity present
in the code, only the usual artificial viscosity to model the presence
of shocks.

Initial conditions are generated by distributing $N$ fluid elements,
over the torus volume after specifying values for $M$, $l(r)$ and
$\Phi_{0}$, and relaxing them for several dynamical times in order to
obtain a distribution close to equilibrium (see Figure~2). This
configuration is then evolved in time with the desired perturbation to
study its dynamical behaviour.

Our initial configurations are thus non--accreting tori fully
contained within their Roche lobe. Specifically, the inner edge of the
disc and the inner Lagrange point, $L_{1}$ are located at
$r_{in}=2.60r_g$ and $r_{L_{1}}=2.25r_g$ respectively. During the
dynamical evolution described below, no accretion takes place, so the
potential produced by the black hole is unaltered.

The analysis of the data is carrried out by performing the Fourier
decomposition of the main hydrodynamic variables as functions of time
(e.g., the position of the center of the disc, the maximum and mean
densities, and the various total energies).

\begin{figure}
\includegraphics[width=84mm]{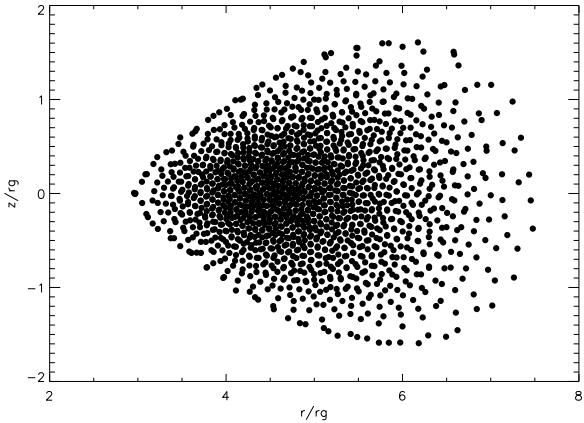}
\caption{An example of a thick torus in hydrodynamical equilibrium.
This snapshot shows the relaxed initial conditions.  Each point
represents a gas cell.  The locus of maximum density is at $r = 4.25
r_g$.}
\label{fig2}
\end{figure}

\section{INTRODUCING A PERTURBATION}

The perturbation acting on the disc can be considered as arising from
the central object, through its magnetic field, a deformation on its
surface or a changing radiation field (in the case of neutron stars)
or as the emission of gravitational waves from an accreting black
hole. Instabilities in the accretion disc itself are another possible
source of time variability, which can induce oscillations in the
fluid. In either case, these would presumably be more intense at small
radii. If the spin of the central object is involved in producing the
perturbation, its amplitude will vary and repeat at intervals given by
the inverse of the spin period, $\Delta T=1/\nu_s$.

We have chosen, then, a perturbation which induces an acceleration in
the disc given by:
\begin{equation}
a_{pert}=-\eta \, a_{g} \cdot \exp\left(\frac{r_o-r}{\delta r}\right)
\cdot \sin\left(2 \pi \nu_s t \right).
\end{equation}
Here $a_{g}$ is the acceleration due to gravity, $r_o$ is the outer
edge of the torus and $\eta \ll 1$ is a parameter that modulates the
strength of the perturbation. The exponential term decays on a scale
$\delta r \simeq R$, the radial extension of the disc, thus
reproducing the desired behaviour for this perturbative force, which
will be strong near the inner radius and weak in the outer regions.
This acceleration induces radial oscillations in the disc, which can
be Fourier--analyzed to extract the main frequencies as was done by
\citet{lee2} recently for a slender torus.

\section{RESULTS AND DISCUSSION}
\begin{figure}
\includegraphics[width=84mm]{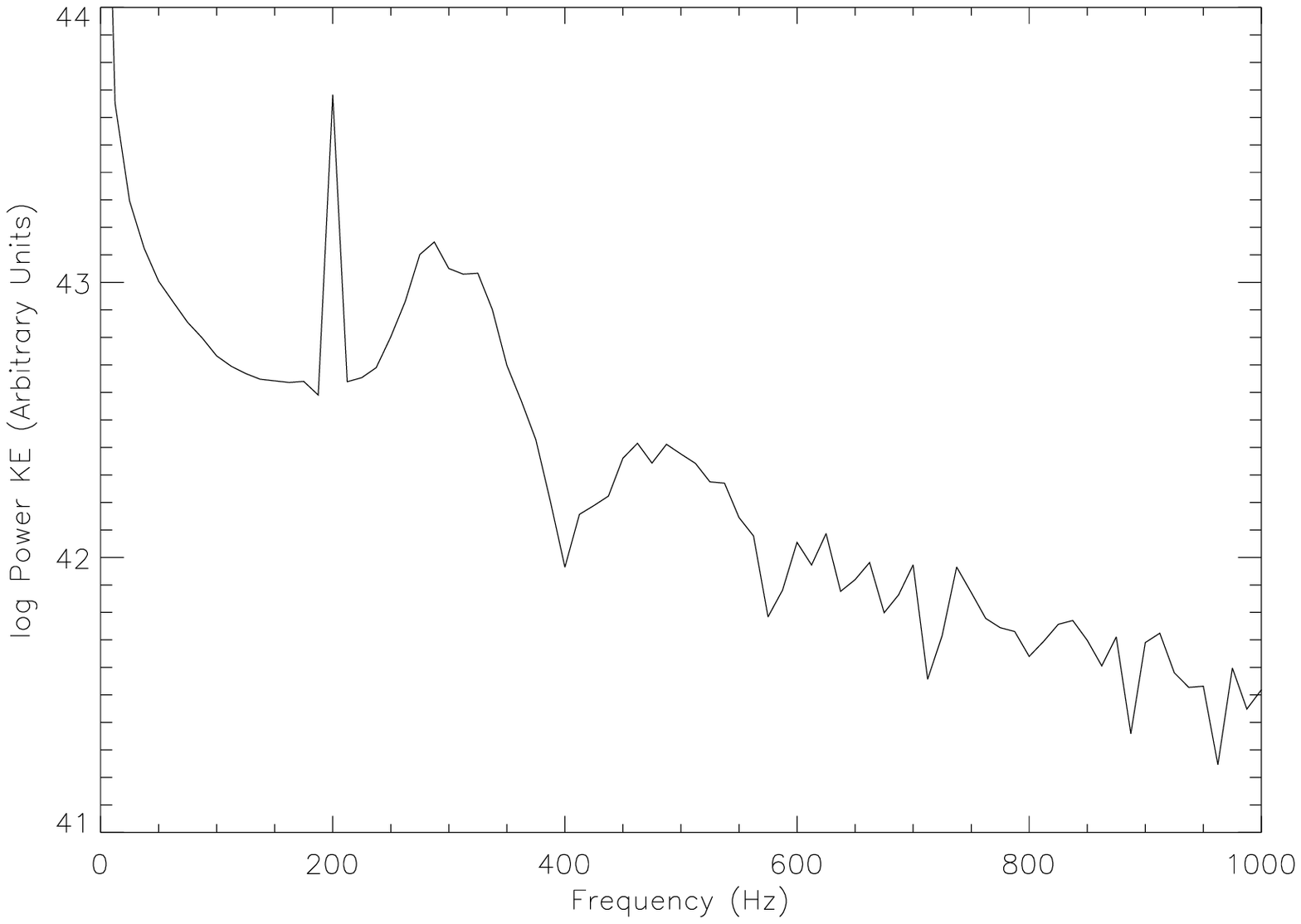} \\

\includegraphics[width=84mm]{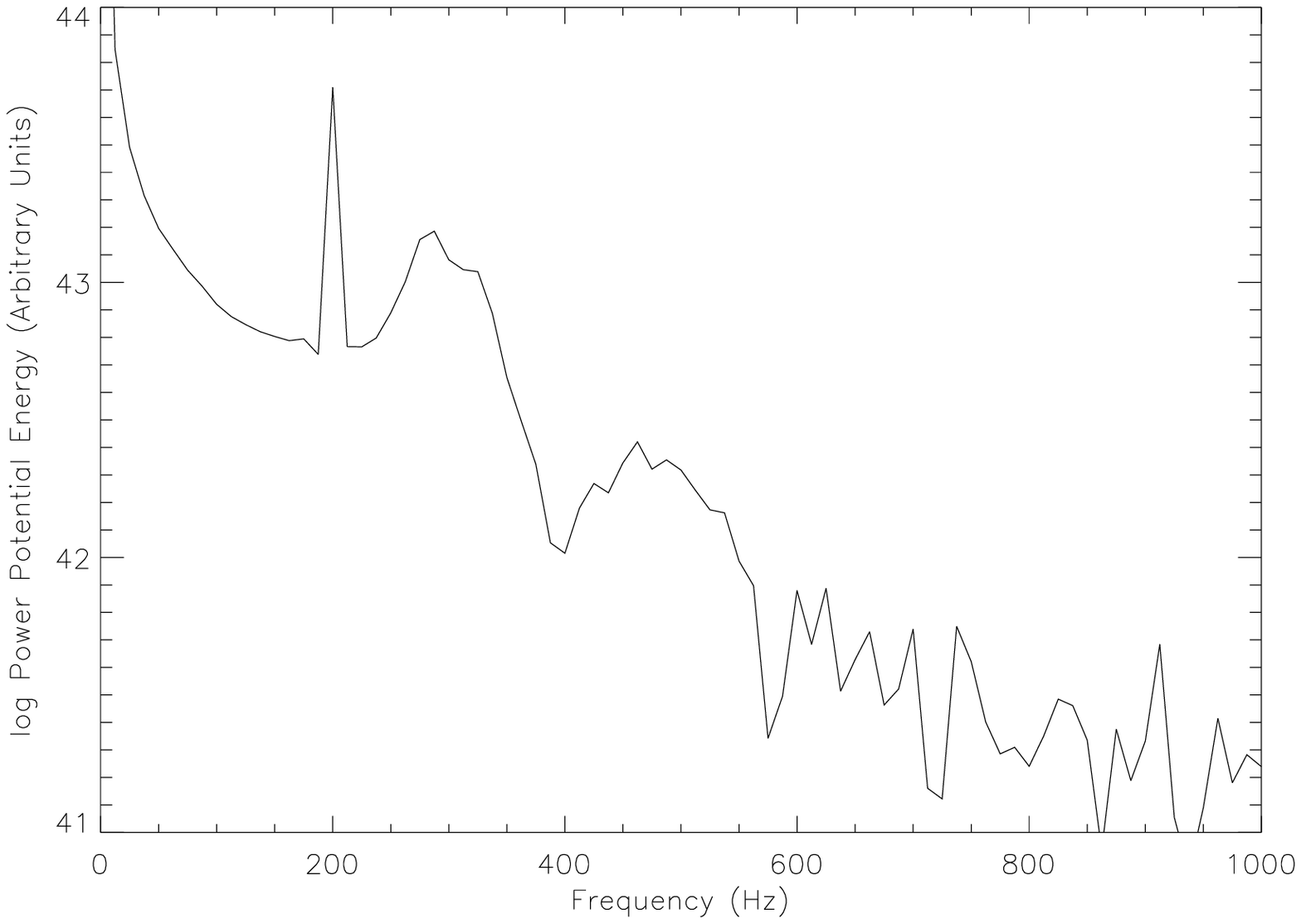} 
\caption{FFT spectra for the kinetic (top) and total potential energy
(bottom), for a torus orbiting a 2.5$M_{\odot}$ black hole.  From left
to right, the first peak corresponds to the perturbation introduced at
200 Hz, the second to the radial epicyclic frequency, shifted to lower
frequecies due to the size of the torus (the first acoustic p--mode),
and the third to 1.5 times the second.}
\label{fig3}
\end{figure}

A Fourier transform of the total kinetic energy in the torus in a
typical calculation is shown in Figure~3. The black hole mass in this
case was 2.5 solar masses, and the perturbation frequency was fixed at
$\nu_{s}=200$~Hz, which clearly shows up prominently in the spectrum 
as a narrow peak. 

Two additional broad features are clearly seen, centered at
$\nu_{1}\approx 300$~Hz and $\nu_{2}\approx 450$~Hz, and are, to the
limit of our resolution, in a 3:2 ratio. The radial epicyclic
frequency for a test particle at the locus of maximum density ($4.25
r_{g}$) is in this case $\kappa=426$~Hz (see equation
\ref{repi}). Trial runs with tori of different radial extent
(Rubio-Herrera, in preparation) and a comparison with the work of
\citet{za} and \citet{rez03a} show clearly that the lower of the two
peaks is simply the epicyclic frequency at the locus of maximum
density, shifted to lower values because of the finite extent of the
torus, i.e. the fundamental acoustic p--mode. The second peak at
higher frequency is the second in the sequence 2:3:4:... of acoustic
p--modes.

We draw from these results the following conclusions. 
\begin{itemize}
\item Global modes of oscillation in thick tori, as studied in the
relativistic regime \citep{za}, can be efficiently excited by a
localized perturbation, affecting only the inner edge of the torus
strongly. The characteristic frequencies apparent in the Fourier
decomposition of the total internal energy exhibit the sequence
2:3:... .
\item The lower of the two frequencies in the series is the first
acoustic p--mode, closely tied to the radial epiclyclic frequency for
test particles, and shifted to lower frequencies because of the finite
extent of the torus.
\end{itemize}

The nature of the excitation mechanism, as stated above, is left as an
open question. It could be some disturbance associated with the pulsar
spin frequency in the case of neutron star systems, or oscillations in
the disc itself which excite these modes. In either case, the
modulation in the X--ray lightcurve will presumably occur in the
innermost regions of the accretion flow. Since its inner boundary is
defined most likely by the effects of strong gravity, one would expect
that the frequencies would scale inversely with the mass of the
central object, as is indeed the case \citep{mcclin} for X--ray
binaries --- and may in principle allow for a mass determination in
the intermediate mass black hole candidates \citep{akmr04}.

Two potentially limiting simplifications in this study deserve
justification. First, in the context of LMXBs, the ratio of disc mass
to black hole mass is very low, $M_d/M_{BH} \ll 1$.  Hence it is
reasonable to suppose that the gravitational potential of the disc is
negligible when compared to that of the black hole, as we have
assumed. Second, we have not considered the effects of magnetic
fields. This is simply because we wish to carry out a study of global,
purely hydrodynamical modes, and does not imply that MHD effects are
negligible or irrelevant. Recently, \citet{k04} has obtained
interesting results in the context of kHz QPOs through
three--dimensional MHD calculations performed in a pseudo--Newtonian
potential.

Finally, one may question the choice of a constant distribution of
specific angular momentum within the torus, assumed here for
simplicity and as a first step. However, we note that in viscous,
hydrodynamical flows, even if the injection of matter at large
distances occurs at nearly Keplerian values of the angular momentum,
the flow near the compact object may follow a flatter distribution
\citep{ia99}. The Papaloizou--Pringle instability \citep{pp84} appears
when non--axisymmetric perturbations act on non--accreting discs with
constant angular momentum.  When accretion is taken into account, this
instability is suppressed, as was shown by \citet{b87}.  This leads to
flat distributions of angular momentum near the central object and a
power law distribution in the outer region of the torus. We thus
believe it is reasonable to assume a constant angular momentum for the
discs as a first approximation.

How the oscillations of the fluid in the disc may translate into
variations in the X--ray lightcurve and be observed as kHz QPOs is
ultimately still unsolved, and requires more complex physical
processes than those included here. Assuming that oscillations at such
frequencies do in fact occur due to the presence of inhomogeneities in
the accretion flow, the luminosity modulation with account of
spacetime curvature in the vicinity of a black hole has been
investigated by \citet{sch04a} and \citet{sch04b}. Their results
indicate that a range of variation would in fact be reflected in the
X--rays, which encourages the investigation of simple modes of fluid
oscillation. In future work we will discuss the behaviour of discs
with non--constant angular momentum subject to various perturbations.

\section*{ACKNOWLEDGMENTS}

It is a pleasure to acknowledge many suggestions and discussions
on this matter with L. Rezzolla. Financial support for this work 
was provided by CONACyT (36632E).

 \bsp  

\label{lastpage}


\vspace{-0.3cm}
\begin{thebibliography}{}
\bibitem[\protect\citeauthoryear{Abramowicz, Bulik, Bursa \&
    Klu\'zniak}{2003}]{ab3} Abramowicz, M.A., Bulik, T., Bursa, M.,
    Klu\'zniak, W.  2003, A\&A, 404, L21
\bibitem[\protect\citeauthoryear{Abramowicz, Karas, Klu\'{z}niak, Lee
    \& Rebusco}{2003}]{ab} Abramowicz, M.A., Karas, V., Klu\'{z}niak,
    W., Lee, W. H., Rebusco, P. 2003, PASJ, 55, 467
\bibitem[\protect\citeauthoryear{Abramowicz \&
    Klu\'{z}niak}{2001}]{ab2} Abramowicz, M.A., Klu\'{z}niak, W. 2001,
    A\&A, 374, L19
\bibitem[\protect\citeauthoryear{Abramowicz, Calvani \& Nobili}{1983}]{ab4} 
    Abramowicz, M.A., Calvani, M., Nobili, L. 1983, Nature, 302, 597
\bibitem[\protect\citeauthoryear{Abramowicz, Klu\'{z}niak, McClintock
    \& Remillard}{2004}]{akmr04} Abramowicz, M.A., Klu\'{z}niak, W.,
    McClintock, J.E., Remillard, R. 2004, ApJ, 609, L63
\bibitem[\protect\citeauthoryear{Blaes}{1987}]{b87} 
     Blaes, O.M., 1987, MNRAS, 227, 975
\bibitem[\protect\citeauthoryear{Daigne \& Mochkovitch}{1997}]{dm97} 
     Daigne, F., Mochkovitch, R., 1997, MNRAS, 285, L15
\bibitem[\protect\citeauthoryear{Igumenshchev \&
   Abramowicz}{1999}]{ia99} Igumenshchev, I.V., Abramowicz, M.A. 1999,
   MNRAS, 303, 309
\bibitem[\protect\citeauthoryear{Kato}{2004}]{k04} 
   Kato, Y., 2004, PASJ, 56, 931
\bibitem[\protect\citeauthoryear{Klu\'zniak, Michelson \&
     Wagoner}{1990}]{kl} Klu\'zniak, W., Michelson, P., Wagoner, R.
     1990, ApJ, 358, 538
\bibitem[\protect\citeauthoryear{Klu\'{z}niak, Abramowicz, Kato, Lee
    \& Stergioulas}{2004}]{kakls04} Klu\'zniak, W., Abramowicz, M.A.,
    Kato, S., Lee, W.H., Stergioulas, N. 2004, ApJ, 603, L89
\bibitem[\protect\citeauthoryear{Lee}{2001}]{l01}
    Lee, W.H. 2001, MNRAS, 328, 583 
\bibitem[\protect\citeauthoryear{Lee \& Ramirez--Ruiz}{2002}]{lee3}
    Lee, W.H., Ramirez-Ruiz, E. 2002, ApJ, 577, 893
\bibitem[\protect\citeauthoryear{Lee, Abramowicz \& Klu\'zniak}{2004}]{lee2}
    Lee, W.H., Abramowicz, M.A., Klu\'zniak, W. 2004, ApJ, 603, L93
\bibitem[\protect\citeauthoryear{McClintock \&
   Remillard}{2004}]{mcclin} McClintock, J.E., Remillard, R. 2004, in
   Compact Stellar X-Ray Sources, ed. W. H. G. Lewin \& M. van der Klis
   (Cambridge: Cambridge Univ. Press), in press (astro-ph/0306213)
\bibitem[\protect\citeauthoryear{Mineshige, Hosokawa, Mashida \&
  Matsumoto}{2002}]{mi} Mineshige, S., Hosokawa, T., Machida, M.,
  Matsumoto, R. 2002 PASJ, 54, 655
\bibitem[\protect\citeauthoryear{Monaghan}{1992}]{mo}
   Monaghan, J. 1992, ARA\&A, 30, 543
\bibitem[\protect\citeauthoryear{Montero, Rezzolla \&
   Yoshida}{2004}]{mry04} Montero, P., Rezzolla, L., Yoshida, S. 2004,
   MNRAS, 354, 1040
\bibitem[\protect\citeauthoryear{Narayan, Paczy\'{n}ski \&
   Piran}{1992}]{npp} Narayan, R., Paczy\'{n}ski, B., Piran, T.  1992,
   ApJ, 395, L83
\bibitem[\protect\citeauthoryear{Paczy\'nski \& Wiita}{1980}]{pw}
   Paczy\'nski, B., Wiita, J. 1980, A\&A, 88, 23
\bibitem[\protect\citeauthoryear{Papaloizou \& Pringle}{1984}]{pp84}
   Papaloizou, J. C. B.; Pringle, J. E. 1984, MNRAS, 208, 721
\bibitem[\protect\citeauthoryear{Rebusco}{2004}]{rebusco04}
   Rebusco, P. 2004, PASJ, 56, 553
\bibitem[\protect\citeauthoryear{Rezzolla, Yoshida \& Zanotti}{2003}]{rez03a}
   Rezzolla, L., Yoshida, S., Zanotti, O. 2003 MNRAS 344, 978
\bibitem[\protect\citeauthoryear{Rezzolla, Yoshida, Maccarone \&
   Zanotti}{2003}]{rez03b} Rezzolla, L., Yoshida, S., Maccarone, T.J.,
   Zanotti, O. 2003 MNRAS 344, L37
\bibitem[\protect\citeauthoryear{Ruffert \&
Janka}{1999}]{rj99}Ruffert, M., Janka, H.-Th. 1999, A\&A, 344, 573
\bibitem[\protect\citeauthoryear{Schnittman \& Bertschinger}{2004}]{sch04a}
    Schnittman, J.D., Bertschinger, E. 2004, ApJ, 606, 1098
\bibitem[\protect\citeauthoryear{Schnittman}{2004}]{sch04b}
    Schnittman, J.D. 2004, ApJ submitted (astro-ph/0407179)
\bibitem[\protect\citeauthoryear{van der Klis}{2000}]{vdk00} van der
    Klis, M. 2000, ARA\&A, 38, 717
\bibitem[\protect\citeauthoryear{van Putten}{2001}]{vp}
    van Putten, M.H.P.M. 2001, Phys. Rev. Lett. 87, 091101
\bibitem[\protect\citeauthoryear{Wagoner, Silbergleit \&
 Ortega-Rodr\'{\i}guez}{2001}]{wa2} Wagoner, R., Silbergleit, A.,
 Ortega-Rodr\'{\i}guez, M. 2001, ApJ, 559, L25
\bibitem[\protect\citeauthoryear{Wijnands, van der Klis, Homan,
 Chakrabarty, Markwardt \& Morgan}{2003}]{wi} Wijnands, R., van der
 Klis, M., Homan, J., Chakrabarty, D., Markwardt, C., Morgan, E.,
 2003, Nat., 424, 44
\bibitem[\protect\citeauthoryear{Woosley}{1993}]{wo} Woosley,
S.E. 1993, ApJ 405, 273
\bibitem[\protect\citeauthoryear{Zanotti, Rezzolla \& Font}{2003}]{za}
 Zanotti, O., Rezzolla, L., Font, J. 2003 MNRAS 341, 832
\bibitem[\protect\citeauthoryear{Zurek \& G\'orski}{1989}]{zu}
    Zurek, W., G\'orski, K. 1989 ApJ, 347, L47
\end{thebibliography}
\end{document}